\documentclass[aps,twocolumn,showpacs,footinbib]{revtex4}%
\usepackage[latin9]{inputenc}
\usepackage{amssymb}
\usepackage{amsmath}
\usepackage{amscd}
\usepackage{graphicx}
\usepackage{subfigure}
\usepackage{float}

\begin{document}
\draft
\title{Dicke-like quantum phase transition and vacuum entanglement with two coupled
atomic ensembles}
\author{Shi-Biao Zheng\thanks{%
E-mail: sbzheng@pub5.fz.fj.cn}}
\address{Department of Physics\\
Fuzhou University\\
Fuzhou 350002, P. R. China}
\date{\today }

\begin{abstract}
We study the coherent cooperative phenomena of the system composed of two
interacting atomic ensembles in the thermodynamic limit. Remarkably, the
system exhibits the Dicke-like quantum phase transition and entanglement
behavior although the governing Hamiltonian is fundamentally different from
the spin-boson Dicke Hamiltonian, offering the opportunity for investigating
collective matter-light dynamics with pure matter waves. The model can be
realized with two Bose-Einstein condensates or atomic ensembles trapped in
two optical cavities coupled to each other. The interaction between the two
separate samples is induced by virtual photon exchange.
\end{abstract}

\maketitle \vskip 0.5cm

\narrowtext

\section{INTRODUCTION}

Entanglement, arising from nonlocal superposition state of two or more
quantum systems, is one of the most striking features of quantum mechanics
and plays a key role in the test of local hidden variable theories [1,2].
Over the past two decades entanglement is regarded as a key resource to
implement quantum information tasks, such as quantum cryptography [3],
computer [4], and teleportation [5]. On the other hand, entanglement is
closely related to various collective quantum phenomena and plays a central
role in studying quantum many-body systems. Typical examples are entangled
ground states used to clarify superconductivity [6] and fractional quantum
Hall effect [7]. Ground state entanglement, which is responsible for
long-range correlations at zero temperature [8], may shed light upon quantum
phase transition [8]. The connection between entanglement and quantum phase
transition has been extensively explored in the quantum system made up of N
spins 1/2 on a one-dimensional lattice [9-11] or on a simplex [12,13].

Another important example of correlated many-particle systems is the Dicke
model [14], which describes the interaction of an ensemble of N two-level
systems with a single-mode quantized field. When the atom-field coupling is
strong enough such a model exhibits a superradiant quantum phase transition
[15-17]. Lambert et al. have investigated the critical behavior of the
atom-field entanglement and pairwise entanglement between atoms in the
thermodynamics limit [18]. The Dicke model has also been a fertile ground
for implementation of quantum information as atomic ensembles with
long-lived electronic states are ideal for storing and processing local
quantum information via interaction with light fields. High-fidelity quantum
operations and entanglement may be achieved beyond the strong coupling
between a single atom and a single photon due to the collective enhancement
of interaction with the field mode [19].

The Dicke quantum phase transition occurs under the condition that the
coupling strength is on the order of the energy separation between two
involved levels and thus the counter-rotating terms significantly affect the
dynamics, which is difficult to achieve in typical cavity QED systems since
the atomic transition frequencies usually exeed the atom-cavity coupling
strength by many orders of magnitude. Dimer et al. have propsed an effective
Dicke model based on balanced Raman transitions between two ground atomic
states of an atomic ensemble induced by a cavity mode and a pair of laser
fields [20], which significantly lowers the energy difference between the
two involved levels. It has been shown that the Dicke model can also been
realized in a laser-driven Bose-Einstein condensate coupled to an optical
cavity and the onset of self-organization corresponds to the Dicke quantum
phase transition [21-23]. However, these systems are subject to cavity loss,
which would significantly alter the coherent dynamics of the Hamiltonian and
deteriorate the matter-light entanglement. Though the superradiant quantum
phase transition has been observed in Ref. [22], the ground state
entanglement has not been verified.

In this paper we investigate the collective dynamics of the system that
consists of two coupled atomic ensembles and find two main results. Firstly,
we show that the coupled spin system can display the quantum phase
transition and vacuum entanglement properties of the physically distinct
spin-boson Dicke model in the thermodynamic limit, providing an access to
the intriguing critical entanglement of the Dicke model without involving
the dynamics of the quantized field. Secondly, as an example for the
physical implementation of this model we demonstrate that two Bose-Einstein
condensates (BECs) or atomic ensembles trapped in two coupled optical
cavities can form the effective spin Hamiltonian. With the assistance of
suitable chosen external fields, the virtual excitation of the cavity modes
mediates effective coupling between the two spin ensembles. Besides
fundamental interest, the system offers the possibility to produce two-mode
squeezed vacuum states between two spatially separated macroscopic systems,
which are of crucial importance for quantum communication [24] and
nonlocality test [25] with continuous variables. Such states are also useful
for exploring the boundary between quantum and classical worlds and
understanding the decoherence effect in quantum information.

The paper is organized as follows. In Sec.2, we study the Hamiltonian
dynamics of two coupled atomic ensembles in the thermodynamic limit, and
show that the system can exhibit the quantum phase transition and critical
entanglement behavior of the Dicke model. In Sec.3, we propose an
experimental realization of the system by coupling the motional degrees of
freedom of two BECs trapped in two coupled optical cavities. The model can
also be realized in the electronic degrees of freedom of two atomic
ensembles via virtual photon exchange. A summary appears in Sec.4.

\section{QUANTUM PHASE TRANSITION AND VACUUM ENTANGLEMENT}

Let us start by considering the system involving two coupled atomic
ensembles, with the $i$th ($i=1$, $2$) ensemble consisting of $N_i$
two-level atoms of energy splitting $\omega _i$. The Hamiltonian of the
system is given by

\begin{equation}
H=\omega _1J_{1,z}+\omega _2J_{2,z}+\frac \lambda {\sqrt{N_1N_2}}%
(J_1^{+}+J_1^{-})(J_2^{+}+J_2^{-}),
\end{equation}
where \{$J_{i,z},J_i^{\pm }\}$ are collective atomic operators for the $i$th
ensemble, satisfying the angular momentum commutation relations $%
[J_i^{+},J_i^{-}]=2J_{i,z}$ and $[J_i^{\pm },J_{i,z}]=\mp J_i^{\pm }$. The
Hamiltonian commutes with the partity operator $e^{i\pi (J_{1,z}+J_{2,z})}$.
Using the Holstein-Primakoff representation, we express the operators \{$%
J_{i,z},J_i^{\pm }\}$ in terms of the annihilation and creation operators $%
b_i$ and $b_i^{\dagger }$ of a bosonic mode via $J_i^{+}=b_i^{\dagger }\sqrt{%
N_i-b_i^{\dagger }b_i}$, $J_i^{-}=\sqrt{N_i-b_i^{\dagger }b_i}b_i$, and $%
J_{i,z}=b_i^{\dagger }b_i-N_i/2$. Then we obtain the two-mode bosonic
Hamiltonian
\begin{eqnarray}
H &=&\omega _1b_1^{\dagger }b_1+\omega _2b_2^{\dagger }b_2\cr&&+\frac \lambda {%
\sqrt{N_1N_2}}\left( b_1^{\dagger }\sqrt{N_1-b_1^{\dagger }b_1}+\sqrt{%
N_1-b_1^{\dagger }b_1}b_1\right) \cr &&\times\left( b_2^{\dagger
}\sqrt{N_2-b_2^{\dagger }b_2}+\sqrt{N_2-b_2^{\dagger }b_2}b_2\right)
.
\end{eqnarray}
Perform the transformation $D_1^{+}(\alpha )D_2^{+}\left( \beta \right)
HD_1(\alpha )D_2\left( \beta \right) $, where $D_1(\alpha )=e^{\alpha
b_i^{\dagger }-\alpha ^{*}b_i}$ and $D_2(\beta )=e^{\beta b_i^{\dagger
}-\beta ^{*}b_i}$ are displacement operators. In the thermodynamic limit $%
N_i\rightarrow \infty $ we can approximate the square root terms up to $%
1/N_i $. The resulting Hamiltonian is expanded up to the second-order in the
boson operators.

In order for the coefficients of the linear terms in the displaced
Hamiltonian to be zero, the displacement amounts $\alpha $ and $\beta $
should satisfy
\begin{eqnarray}
\omega _1\alpha +2\lambda \beta \frac{N_1-2\alpha ^2}{\sqrt{N_1N_2}}\sqrt{%
\frac{N_2-\beta ^2}{N_1-\alpha ^2}} &=&0, \cr
\omega _2\beta +2\lambda \alpha \frac{N_2-2\beta ^2}{\sqrt{N_1N_2}}\sqrt{%
\frac{N_1-\alpha ^2}{N_2-\beta ^2}} &=&0.
\end{eqnarray}
Under the critical point $\lambda _c=\sqrt{\omega _1\omega _2}/2$, the
solution is $\alpha =\beta =0$, which corresponds to the normal phase. In
this case the effective Hamiltonian $H^{(1)}=\omega _1b_1^{\dagger
}b_1+\omega _2b_2^{\dagger }b_2+\lambda (b_1^{\dagger }+b_1)(b_2^{\dagger
}+b_2)$ is mathematically equivalent to the spin-boson Dicke Hamiltonian in
the normal phase [15]. The ground state in this phase is a two-mode squeezed
vacuum state, which has a definite partiy and is incoherent, i.e., $%
\left\langle (J_i^{+}+J_i^{-})\right\rangle =0$.

Above the critical point, there exist two physically sensible solutions \{$%
\alpha =\sqrt{N_1}\alpha _0,\beta =-\sqrt{N_2}\beta _0$\} and \{$\alpha =-%
\sqrt{N_1}\alpha _0$, $\beta =\sqrt{N_2}\beta _0$\}, where $\alpha _0=\{%
\frac 12[1-\sqrt{\frac{4\lambda ^2\omega _1^2+r(\omega _1\omega _2)^2}{%
4\lambda ^2\omega _1^2+16r\lambda ^4}}]\}^{1/2}$, $\beta _0=\{\frac 12[1-%
\sqrt{\frac{4r\lambda ^2\omega _2^2+(\omega _1\omega _2)^2}{4r\lambda
^2\omega _2^2+16\lambda ^4}}]\}^{1/2}$, and $r=N_2/N_1$. The corresponding
effective Hamiltonian is
\begin{eqnarray}
H^{(2)} &=&\frac{K_1+K_2}2b_1^{\dagger
}b_1+\frac{K_1-K_2}4(b_1^{\dagger 2}+b_1^2) \cr
&&+\frac{K_3+K_4}2b_2^{\dagger }b_2+\frac{K_3-K_4}4(b_2^{\dagger
2}+b_2^2)\cr&&+K_5(b_1^{\dagger }+b_1)(b_2^{\dagger }+b_2),
\end{eqnarray}
where
\begin{eqnarray}
K_1 &=&\omega _1-\frac{2\alpha _0\beta _0(3-2\alpha
_0^2)\sqrt{r(1-\beta _0^2)}\lambda }{(1-\alpha _0^2)^{3/2}}, \cr
K_2 &=&\omega _1-2\alpha _0\beta _0\lambda \sqrt{\frac{r(1-\beta _0^2)}{%
1-\alpha _0^2}},  \nonumber \\
K_3 &=&\omega _2-\frac{2\alpha _0\beta _0(3-2\beta _0^2)\sqrt{(1-\alpha
_0^2)/r}\lambda }{(1-\beta _0^2)^{3/2}},  \nonumber \\
K_4 &=&\omega _2-2\alpha _0\beta _0\lambda \sqrt{\frac{1-\alpha _0^2}{%
r(1-\beta _0^2)}},  \cr K_5 &=&\frac{(1-2\alpha _0^2)(1-2\beta
_0^2)\lambda }{(1-\alpha _0^2)^{1/2}(1-\beta _0^2)^{1/2}}.
\end{eqnarray}
The Hamiltonian $H^{(2)}$ is diagonal in terms of two normal bosonic modes
with the frequencies

\begin{eqnarray}
\omega _{\pm }^2&=&\frac{K_1K_2+K_3K_4}2\cr&&\pm \left[ \frac{(K_1K_2-K_3K_4)^2}4%
+4K_2K_4K_5^2\right] ^{1/2}.
\end{eqnarray}
In the regime near the critical point, the energy gap to the first
excited state vanishes as $\omega _{-}\propto \left| \lambda
-\lambda _c\right| ^{1/2}$. Above the threshold, in addition to
two-mode squeezing both modes exhibit single-mode squeezing since
the coefficient of $b_i^{\dagger 2}+b_i^2 $ does not vanish for each
mode. However, the single-mode squeezing does not affect the
entanglement between the two collective atomic modes. Below the
threshold, the single-mode squeezing vanishes and thus the threshold
is not affected.

Above the threshould, there are two degenerate ground states, which
corresponds to the breaking of the parity symmetry with the atomic
polarizations $\left\langle (J_1^{+}+J_1^{-})\right\rangle =2\alpha \sqrt{%
N_1-\alpha ^2}$ and $\left\langle (J_2^{+}+J_2^{-})\right\rangle =2\beta
\sqrt{N_2-\beta ^2}$ acquiring macroscopic populations. In this phase both
the coherent excitation and squeezing contribute to the excited atomic
numbers. This is analogous to the superradiant phase of the spin-boson Dicke
model with the field mode replaced by another atomic mode. The excitation
number of each ensemble as a function of $\lambda /\lambda _c$ is ploted in
Fig. 1. The solid line represents the incoherent excitation number due to
squeezing, while the dashed line represents the scaled coherent excitation
number. Near the critical point the incoherent excitation number of each
atomic ensemble due to squeezing is $\stackrel{-}{n}_{inc}\infty \left|
\lambda -\lambda _c\right| ^{-1/4}$, which diverges with the same exponent $%
1/4$ as the correlation length $\xi =\omega _{-}^{-1/2}$. The entanglement
between the two atomic ensembles can be determined by the von Neumann
entropy $S=-$tr$\rho \log _2\rho $, where $\rho $ is the reduced density
operator of one atomic ensemble. In the superexcitation phase, this entropy
is given by

\begin{equation}
S=(k+1/2)\log _2(k+1/2)-(k-1/2)\log _2(k-1/2)+1,
\end{equation}
where $k=\frac 12\left[ 1+\frac{\sin ^22\theta }4(\sqrt{\omega _{+}/\omega
_{-}}-\sqrt{\omega _{-}/\omega _{+}})^2\right] ^{1/2}$ and $\theta =\frac 12%
\arctan [2K_5\sqrt{K_2K_4}/(K_1K_2-K_3K_4)]$. The entanglement as a function
of $\lambda /\lambda _c$ can be seen in Fig. 2. At the critical point the
entanglement diverges logarithmically also with the exponent $1/4$ as $%
S\infty \log _2\left| \lambda -\lambda _c\right| ^{-1/4}$, which is
analogous to the critical behavior of the atom-field entanglement in the
Dicke model [18]. In the above we have assumed that $\lambda >0$. It should
be noted that the system exhibits the same critical behavior and
entanglement for $\lambda <0$.

\section{PHYSICAL REALIZATION}

We note that the Hamiltonian (1) can be realized in the quantum motions of
two BECs or atomic ensembles trapped in two coupled single-mode cavities
along the $x$-axis. The resonant coupling between the two cavity modes is
given by the interaction Hamiltonian $H_c=\nu (a_1^{\dagger }a_2+a_1a_2^{+})$%
, where $a_i^{\dagger }$ and $a_i$ are the creation and annihilation
operators for the $i$th cavity mode, and $\nu $ is the intercavity hopping
strength. Such a coupling can be mediated by overlap of their evanescent
fields or by an optical fiber [26].

We first consider the cavity-BEC system. Suppose that the $i$th BEC is
composed of $N_i$ two-level atoms, each of which is coupled to the $i$th
cavity mode with the maximum coupling strength $g_i$ and driven by a pump
laser field along the $y$-axis with the maximum Rabi frequency $\Omega _i$.
The pump frequency $\omega _p$, close to the cavity mode frequency $\omega _c
$, is highly detuned from the atomic transition frequency $\omega _a$. Then
the atomic excited level can be adiabatically eliminated and the atoms
coherently scatter light between the pump field and the cavity mode, which
induce two balanced Raman channels between the atomic zero-momentum state
and the symmetric superposition of states with the momentum of a photon, $%
\hbar k$, along the $x$ and $y$ directions. Hence each atomic field can be
described in a Hilbert space spanned by two Fourier-modes $c_{0,i}$ and $%
c_{1,i}$, with $c_{0,i}^{\dagger }c_{0,i}+c_{1,i}^{\dagger }c_{1,i}=N_i$
being a constant of motion [21,22]. The coupling between the two modes can
be described by the angular momentum operators: $J_i^{+}=c_{1,i}^{\dagger
}c_{0,i}$, $J_i^{-}=c_{0,i}^{\dagger }c_{1,i}$, and $J_{i,z}=\frac 12%
(c_{1,i}^{\dagger }c_{1,i}-c_{0,i}^{\dagger }c_{0,i})$. In the frame
rotating with the pump field frequency $\omega _p$ the Hamiltonian for the $i
$th BEC-cavity system is
\begin{equation}
H_i=\delta _{c,i}a_i^{\dagger }a_i+\omega _0J_{i,z}-\frac{\eta _i}2%
(a_1+a_1^{\dagger })(J_i^{+}+J_i^{-})-u_ia_i^{\dagger }a_ic_{1,i}^{\dagger
}c_{1,i},
\end{equation}
$\delta _{c,i}=\omega _c-\omega _p-N_ig_i^2/(2\Delta _a)$, $\omega _0=\hbar
k^2/m$, $\eta _i=g_i\Omega _i/\Delta _a$, $u_i=3g_i^2/(4\Delta _a)$, and $%
\Delta _a=\omega _a-\omega _c$. Under the condition $N_1=N_2$ and $g_1=g_2$
we can choose the pump frequency appropriately so that $\delta _{c,i}=0$.
Introducing the new bosonic modes $d_{1,2}=\frac 1{\sqrt{2}}(a_1\pm a_2)$,
we can rewrite the Hamiltonian of the total system as
\begin{eqnarray}
H_t &=&\nu d_1^{\dagger }d_1-\nu d_2^{\dagger }d_2+\omega
_0J_{1,z}+\omega _0J_{2,z} \cr&&\ -\frac{\eta
_1}{2\sqrt{2}}(d_1+d_1^{\dagger }+d_2+d_2^{\dagger
})(J_1^{+}+J_1^{-})  \nonumber \\
&&\ -\frac{\eta _2}{2\sqrt{2}}(d_1+d_1^{\dagger }-d_2-d_2^{\dagger
})(J_2^{+}+J_2^{-})  \nonumber \\
&&\ \ \ \ \ -\frac{u_1}2(d_1^{\dagger }d_1+d_2^{\dagger }d_2+d_1^{\dagger
}d_2+d_2^{\dagger }d_1)c_{1,1}^{\dagger }c_{1,1}  \nonumber \\
&&\ \ \ \ \ -\frac{u_2}2(d_1^{\dagger }d_1+d_2^{\dagger
}d_2-d_1^{\dagger }d_2-d_2^{\dagger }d_1)c_{1,2}^{\dagger }c_{1,2}.
\end{eqnarray}
Under the condition $\nu \gg \sqrt{N_i}\eta _i$, $\omega _0$, $\stackrel{-}{n%
}_iu_i$, $\delta _{c,i}$ , where $\stackrel{-}{n}_i$ being the mean motional
excitation number in the ith BEC, one can adiabatically eliminate the field
modes to obtain the coupling between the two condensates. So the effective
Hamiltonian is given by
\begin{eqnarray}
H_e &=&\nu d_1^{\dagger }d_1-\nu d_2^{\dagger }d_2+\omega
_0J_{1,z}+\omega _0J_{2,z} \cr &&+\frac \lambda
{\sqrt{N_1N_2}}(J_1^{+}+J_1^{-})(J_2^{+}+J_2^{-}) \cr &&\ \ \ -\frac
12(u_1d_1^{\dagger }d_1+u_2d_2^{\dagger }d_2)(c_{1,1}^{\dagger
}c_{1,1}+c_{1,2}^{\dagger }c_{1,2}) \cr &&\ +\frac 1{4\nu
}(d_1^{\dagger }d_1-d_2^{\dagger }d_2)(u_1c_{1,1}^{\dagger
}c_{1,1}-u_2c_{1,2}^{\dagger }c_{1,2})^2,\cr&&
\end{eqnarray}
where $\lambda =-\sqrt{N_1N_2}\eta _1\eta _2/(2\nu )$. The Hamiltonian
describes a four-photon process which is induced by virtual excitation of
the atomic electronic states and field modes. We note that there is no
coupling between atoms belonging to the same ensembles since the detunings
of the two nonlocal field modes $d_1$ and $d_2$ are opposite, which leads to
opposite contributions to the coupling. On the other hand, these two
nonlocal modes equally contribute to the couplings between atoms belonging
to different ensembles because the product of the two Raman transition
coefficients associated with mode $d_1$ is also opposite to that associated
with $d_2$. When the two cavity modes are both initially in the vacuum
state, the two bosonic modes $d_1$ and $d_2$ will approximately remain in
the vacuum state during the evolution since their frequencies are highly
detuned from the pump frequency due to the strong coupling between the two
cavities. In this case the effective Hamiltonian reduces to Eq. (1), with
the effective coupling strength $\lambda $ being controllable by the Rabi
frequencies or detunings of the pump fields. The quantum phase transition
corresponds to the simultaneous self-organization of the two condensates. In
a realistic experiment, the system is a driven and damping one, which will
realize a steady state governed by energy flow from the pump fields into the
cavity fields, rather than a true ground state of the Hamiltonian, similar
to that studied in Ref. [22].

The Hamiltonian (1) can also be realized in the electronic degrees of
freedom of two atomic ensembles trapped in two coupled cavities. The cavity
mode, together with two external fields, can induce balanced off-resonant
Raman transitions between two ground states of each atomic ensemble [20,27].
With appropriate choice of the parameters of the external fields, the field
modes can be adiabatically eleminated and the two atomic ensembles are
coupled via virtual photon exchange. Due to the stability of the atomic
ground states the vacuum entanglement between the two atomic ensembles
should have a long coherence lifetime and can be readily transferred to
light fields.

\section{SUMMARY}

In conclusion, we have investigated theoretically the ground state
properties of the model involving two coupled spin ensembles in the thermal
limit, showing that the model displays the quantum phase transition and
vacuum entanglement described by the Dicke model despite the fundamental
disctinction between these two models. The model can be realized in the
motions of two BECs or in the internal states of two atomic ensembles in two
coupled cavities. The coupling strength between the two spin ensembles can
be tuned via the parameters of external fields, making the system a
promising simulator for this model. The entanglement within each atomic
ensemble and its connection with the quantum phase transition will be
further investigated. Another interesting problem is how the interaction
between atoms belonging to the same ensembles affects the critical behavior
and entanglement.

This work was supported by the National Natural Science Foundation of China
under Grant No. 10974028, the Doctoral Foundation of the Ministry of
Education of China under Grant No. 20093514110009, and the Natural Science
Foundation of Fujian Province under Grant No. 2009J06002.

\begin{figure}[H]
\includegraphics[width=1\columnwidth]{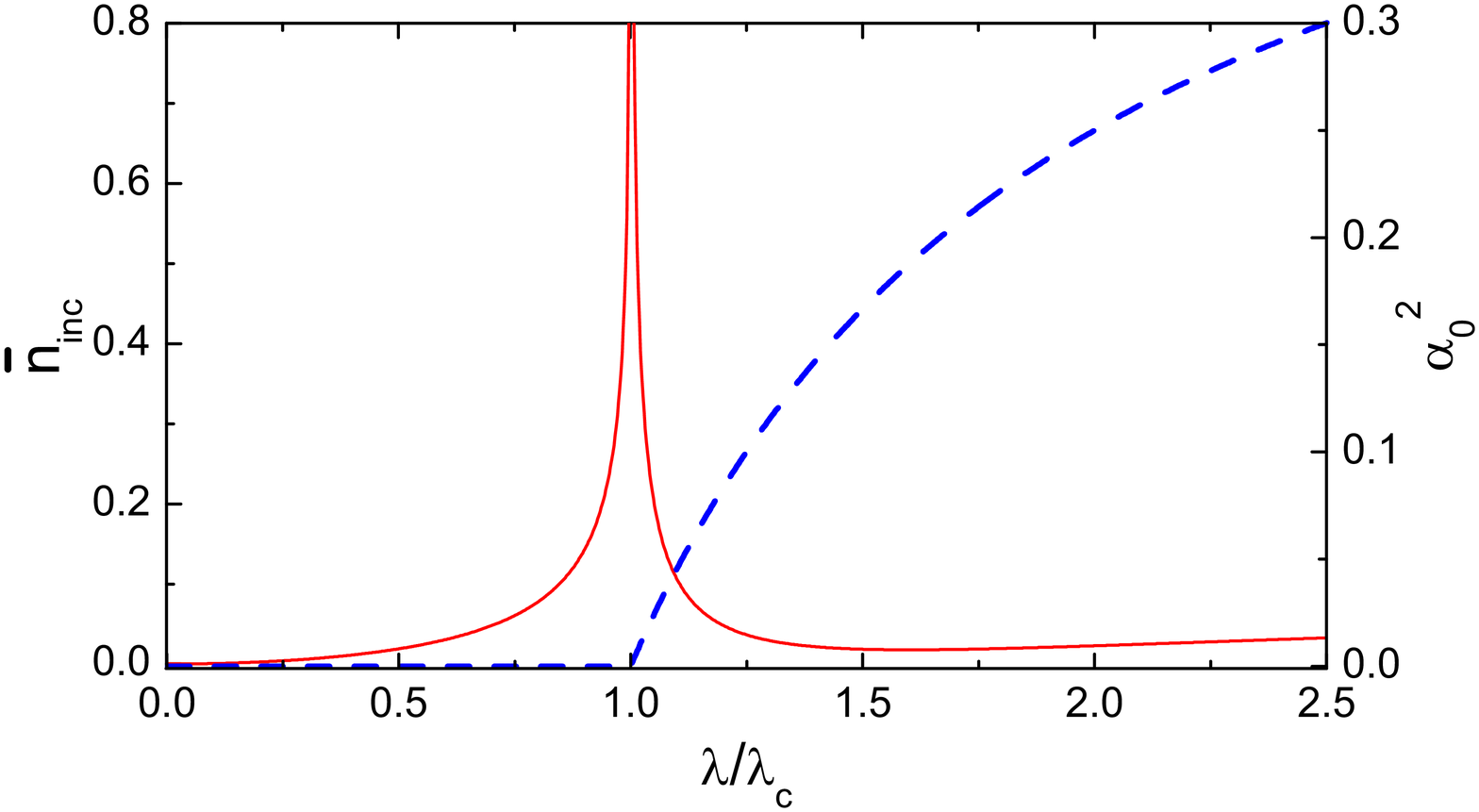}
\caption{(Color online) The mean excitation number as a function of
$\lambda /\lambda _c$. The solid line represents the incoherent
excitation number due to the squeezing, while the dashed line
represents the scaled coherent excitation number, given by the
coherent excitation number devided by $N$. The parameters are
$\omega _1=\omega _2=\omega $, $\lambda _c=\omega /2$, and
$N_1=N_2=N$.}
\end{figure}

\begin{figure}[H]
\includegraphics[width=1\columnwidth]{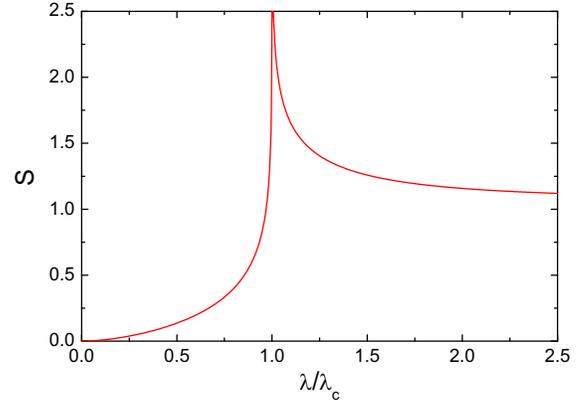}
\caption{(Color online) The entanglement entropy between the two
atomic ensembles as a function of $\lambda /\lambda _c$. The
parameters are $\omega _1=\omega _2=\omega $, $\lambda _c=\omega
/2$, and $N_1=N_2=N$.}
\end{figure}
\end{document}